\documentclass[aps,prl,reprint,superscriptaddress,nofootinbib]{revtex4-2}
\usepackage{xcolor}
\usepackage{mathtools}
\usepackage{amsmath}
\usepackage{amssymb}
\usepackage{graphicx}
\usepackage[unicode=true,pdfusetitle,
bookmarks=true,bookmarksnumbered=false,bookmarksopen=false, breaklinks=true, ,backref=false,colorlinks=true]{hyperref}
\hypersetup{pdfborderstyle=,citecolor=blue}

\makeatletter
\usepackage[normalem]{ulem} 
\usepackage{babel}
\usepackage{array}
\usepackage{multirow}
\usepackage{times}

\makeatother

\begin{document}
\title{Deriving the Forces of Nonequilibria from Two Laws}
\author{Ying-Jen Yang}
\email{ying-jen.yang@stonybrook.edu}
\affiliation{Laufer Center of Physical and Quantitative Biology, Stony Brook University}
\author{Ken A. Dill}
\email{dill@laufercenter.org}
\affiliation{Laufer Center of Physical and Quantitative Biology, Stony Brook University}
\affiliation{Department of Physics, Stony Brook University}
\affiliation{Department of Chemistry, Stony Brook University}
\begin{abstract}

Non-EQuilibrium (NEQ) statistical physics has not had the same general foundation as that of EQuilibrium (EQ) statistical physics, where forces are derived from potentials such as $1/T = \partial S/\partial U$, {and from which other key mathematical relations follow.}  Here, we show {how general NEQ principles can be derived from two corresponding laws.  Maximizing path entropy replaces maximizing state entropy.  Whereas EQ can entail observables $(U, V, N)$, dynamics has (node populations, edge traffic, cycle flux).  We derive forces of NEQ, fluctuation-susceptibility equalities, Maxwell-Onsager-like symmetry relations, and} we generalize to ``cost-benefit'' relations beyond just work and heat dissipation.
\end{abstract}
\maketitle

\textbf{The ``Two Laws'' of nonequilibria:}
It has been noted how sound and general and principled are the laws of EQuilibrium thermodynamics (EQ)---the First and Second Laws, and the consequent relationships that follow from them \cite{einstein_1949}. 
{In contrast, no such similarly principled foundation has existed for nonequilibria (NEQ) \cite{touchette_large_2009}. Important steps have been taken, reviewed elsewhere \cite{pachter_foundations_2023}, but they are incomplete and fall short of the power that the combined First and Second Laws of Thermodynamcs give in corresponding EQ situations.} 
For equilibria, the First Law is a principle of conservation and the Second Law is a variational principle of Maximum Entropy (Max Ent) probabalistic inference over a space of states.  Combined, these two laws define thermodynamic forces, such as $1/T = (\partial S/\partial U)_{V,N}$ driving heat exchange; they give the Maxwell Relations of symmetries among forces and observables; they give the fluctuational bases of susceptibilities and response functions; they give the cost-benefit trade-offs of energy $\mathrm{d}U = \delta q + \delta w$ between heat and work; and, for statistical physics, they give the basic tool---the equilibrium partition function---for modeling how the material properties of macroscopic systems arise from their underlying microscopic components. Here, we derive equivalently principled relationships for {NEQ} in terms of two laws---dynamical conservation and a variational principle of Maximizing Entropy over a space of \emph{pathway}s (Maximum Caliber) \cite{jaynes_minimum_1980,evans_rules_2004,presse_principles_2013,davis_probabilistic_2018,pachter_foundations_2023}. 
 \\ 

{\textbf{The context of prior work in stochastic dynamics.}
As background, we note two existing frameworks for stochastic forces and flows.  First, Master Equations, Markov Models and Fokker-Planck Equations describe First-Law-like conservation of flows of probabilities \cite{gardiner_stochastic_2009,kampen_stochastic_2007}, as Kirchoff's current law does for non-fluctuating steady states, where the sums of inflows to nodes equals the sums of outflows at steady states.  Second, \emph{Stochastic Thermodynamics} (ST) 
has defined a notion of \emph{force,} namely the \emph{cycle affinity} \cite{schnakenberg_network_1976,hill_free_1989,jiang_mathematical_2004,qian_entropy_2016,polettini_conservation_2016,yang_bivectorial_2021}, whose inner product with flow gives (steady-state) entropy production \cite{jiang_mathematical_2004,ge_physical_2010,esposito_three_2010}. This is often justified by how entropy production relates to the free energy dissipation under Local Detailed Balance (LDB) \cite{van_den_broeck_ensemble_2015,seifert_stochastic_2018}, where the dynamics of the reservoirs must be fast enough to stay near equilibrium. While ST is successful in characterizing dynamical irreversibility and fluctuations \cite{jarzynski_equalities_2011,crooks_nonequilibrium_1998,van_den_broeck_ensemble_2015,evans_fluctuation_2002,seifert_stochastic_2018,barato_thermodynamic_2015,horowitz_thermodynamic_2020,peliti_stochastic_2021}, it requires LDB.  And,
entropy production is \textit{not} a general variational potential \cite{vellela_stochastic_2008,martyushev_restrictions_2014}
---its derivative gives cycle affinity \textit{only} near equilibrium (See Sec. I of \cite{SI}). ST lacks the generality and the Second-Law-like variational principle for deriving potentials, forces, and observable coordinates of NEQ. }
  \\

 {Instead, relatively recent work shows that the general variational principle for NEQ is \emph{Maximum Caliber} (Max Cal) \cite{jaynes_minimum_1980,evans_rules_2004,presse_principles_2013,davis_probabilistic_2018,pachter_foundations_2023}, justified by either inference axioms \cite{jaynes_information_1957,shore_axiomatic_1980,presse_nonadditive_2013,caticha_entropy_2021} or Bayesian conditioning in \textit{Large Deviation Theory}\footnote{LDT is a powerful framework for treating rare fluctuations in certain deterministic limit \cite{touchette_large_2009}, e.g. the infinite-data limit \cite{lu_emergence_2022,yang_statistical_2023} or the hydrodynamic limit \cite{bertini_macroscopic_2015}. It can also be used as a \textit{model-update framework} to make a rare event typical by Bayesian conditioning  \cite{csiszar_conditional_1987,chetrite_nonequilibrium_2013}. The latter is equivalent to Max Cal \cite{chetrite_nonequilibrium_2015,chetrite_variational_2015,yang_statistical_2023}.} (LDT) \cite{van_campenhout_maximum_1981,csiszar_conditional_1987,chetrite_nonequilibrium_2013,chetrite_nonequilibrium_2015,chetrite_variational_2015,yang_statistical_2023}.  Max Cal is to dynamics what Maximum Entropy is to statics; it entails maximizing the \emph{path entropy} (rather than the \emph{state entropy}) subject to constraints imposed by Lagrange multipliers, to predict dynamical distributions \cite{presse_principles_2013}. } 
\\

{However, for neither EQ nor NEQ is knowledge of the two foundational laws alone sufficient to achieve the full power of theory.  The two laws of EQ thermodynamics specify fundamental observables, such as $(U,V,N)$, and give their conjugate forces, their fluctuations and susceptibilities, their Maxwell symmetry relations and their Legendre Transforms. Here, in that same spirit, we derive now for NEQ, complete dynamic variable sets, their corresponding path-entropic forces, their Legendre Transforms, their Maxwell-Onsager-like symmetric response relations, and their fluctuation-susceptibility equalities, from two dynamical laws---a conservation law and a variational principle.}\\

 

{Our treatment here is quite general.  It does not require LDB or linearity of force-flow relations, small gradients or other near-equilibrium approximations, or thermal baths or $k_{\text{B}} T$ (Boltzmann's constant multiplied by temperature) as a source of fluctuations.  It does not require that cost-benefit relationships take the form of heat and work.  For example, the present approach can treat vehicle flows on transportation networks, where costs and benefits can include speed, taxes, tolls, payoffs; or traffic disruptions. It can treat flows of ecological species, where costs and benefits entail resource finding, competition, predation, and fitness.  It can treat chemical or biochemical networks, with desirable or undesirable products.  Or, it can treat cyclic devices such as molecular motors and transduction among different forms of energy and matter.  It need not require energetics---as envisioned by Szilard and Mandelbrot \cite{szilard_uber_1929,mandelbrot_derivation_1964}.  All that is required is statistical counting, in distributions and flows.  A major result here is the derivation of non-equilibrium \emph{forces}, which now opens the way toward designing optimal dynamical systems.  For simplicity, our examples here are time-homogeneous Markov jump processes at steady states, but the framework is generalizable to other Markov or non-Markovian processes \cite{SI}.}  \\

\begin{figure}
\begin{centering}
\includegraphics[width=0.9\columnwidth]{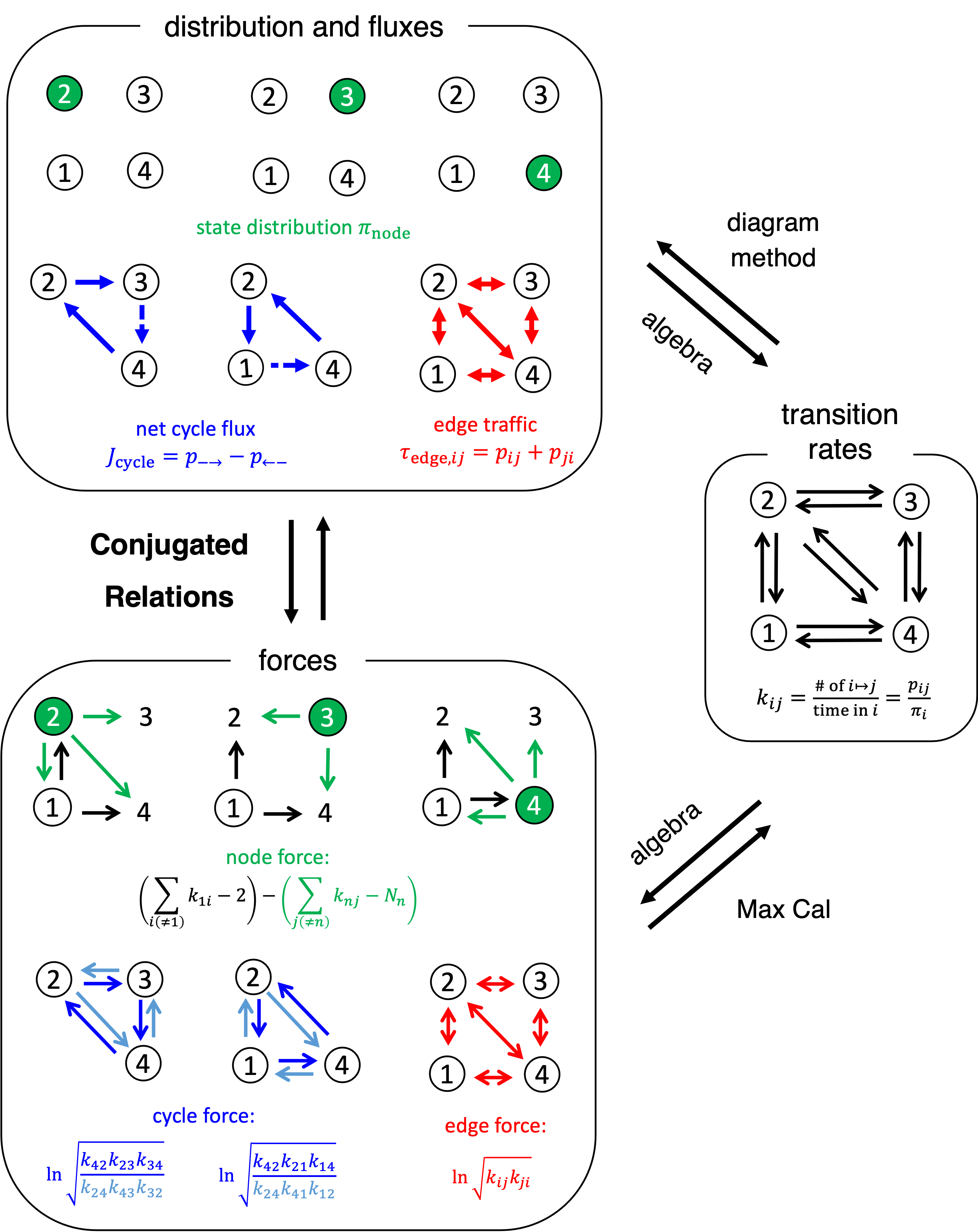}
\par\end{centering}
\caption{ { 
\textbf{Thermodynamics-like observable and force coordinates for the phase space of Markov processes} The phase space of Markov jump processes is often parameterized by the transition rates.
} The upper box shows our {thermodynamic-like observable coordinate} in terms of state distributions, edge traffic flows, and cycle fluxes. The lower box shows the conjugated {coordinate} in terms of the path entropic forces. Arrows between boxes indicate their relations. Rates and forces can be computed from observables with straightforward algebra via their definitions; Observables can be computed from rates by Hill's diagram method; We derive conjugated relations and how to compute rates from forces here.
\label{fig:Parsing}}
\end{figure}

\textbf{The First Law: dynamical conservation.}
Consider a generic network having a state space with nodes $i$ and directed edges $(i,j)$ between the nodes.  The flow is of some kind of agents that are conserved, neither created nor destroyed.  An example is vehicle traffic on a network of roads. The right-hand side of Fig. \ref{fig:Parsing} shows four nodes and five edges.  An agent can be regarded as a vehicle that moves along the routes, capable of staying at the nodes at certain amount of dwell times before moving to another node through an edge. {This can also serve as a simple model for molecular motors like kinesin (cf. Fig. 1c of \cite{liepelt_kinesins_2007}).}  Let $t$ represent the instantaneous time of an agent along a trajectory that runs for a total time of $L$.  Here we suppose $L$ can be arbitrarily long. {We assume the agent follows a Markov jump process with a steady state.  We aim to find a thermodynamics-like observable-based coordinates for its phase space.} \\

{
The elementary counting observables take the form of the probability distribution $\pi_i$ and the fluxes $p_{ij}$ along the edges.  These quantities uniquely determine the transition rates $k_{ij}=p_{ij}/\pi_i$. Now, to get proper coordinates, we must remove their degeneracies that result from \textit{probability conservation}. We first decompose the fluxes $p_{ij}$ into their symmetric and antisymmetric parts $p_{ij} = (\tau_{ij}+J_{ij})/2$.  The symmetric part is  $\tau_{ij}=p_{ij}+p_{ji}$.  This sum of edge fluxes in both directions has been called \textit{traffic} by Maes \textit{et al.} \cite{maes_canonical_2008,maes_frenesy_2020}.  The asymmetric part is the net flux from node $i$ to $j$, $J_{ij}=p_{ij}-p_{ji}$. The degeneracy in $(\pi_i,p_{ij})$ is in normalization $\sum_i \pi_i =1$ and the divergence-free condition on net fluxes $\sum_j J_{ij}=0,\forall i$.
We need non-degenerate parameterization for $\pi_i$ and $J_{ij}$ to form a coordinate.
Hence we leave out one reference node in the coordinate, say node $m$, for the distribution $\pi_i$ and find the linearly-independent components of $J_{ij}$, known as the \textit{fundamental cycle fluxes} $J_c$ \cite{schnakenberg_network_1976,hill_free_1989, cPRE}. A observable coordinate $(\pi_n,\tau_{ij},J_c)$ is thus obtained (upper box in Fig. \ref{fig:Parsing}).} \\

{These observables determine the long-term behavior of generic costs and benefits.}
A trajectory (pathway) {$\omega=i_{0:L}$} of the agent running from time $0$ to $L$, will produce certain time-extensive costs or benefits 
{
\begin{equation} \label{eq: b1 sum and integral}
    \boldsymbol{B}[\omega] =  \int_{0}^{L}\boldsymbol{f}(i_t){\rm d}t + \sum_{t_{n}}\boldsymbol{g}(i_{t_{n-1}},i_{t_{n}})
\end{equation}}where $\boldsymbol{f}(i)$ is the rate of production when the system dwells at node $i$, $\boldsymbol{g}(i,j)$ is production per jump from $i$ to $j$, and $t_n$ is the time immediately after the $n$-th transition. 
{We use boldface variables here to represent a vector of observables.
Below are some examples.  An agent dwelling at node 3 may have lodging costs with a rate $r$: ${f}(i)=r \delta_{i,3}$ and ${g}={0}$ where $\delta_{i,j}$ is the Kronecker delta function. Edge 1-2 may be a bridge collecting tolls $l$ for agent crossing either way: ${f}={0}$ and ${g}(i,j)= l\cdot  (\delta_{i,1}\delta_{j,2}+\delta_{i,2}\delta_{j,1})$. For kinesin, jumping from state 4 to 2 (2 to 4) can correspond to moving one step $d$ forward (backward) in the physical space: ${f}={0}$ and ${g}(i,j)= d\cdot  (\delta_{i,4}\delta_{j,2}-\delta_{i,2}\delta_{j,4})$. 
In the long-trajectory limit,  the production rate of an arbitrary collection of these types of observable $\boldsymbol{B}$ is determined by the steady-state average production rates given by $(\pi_i,p_{ij})$: 
\begin{equation}
    \bar{\boldsymbol{b}} = \lim_{L \rightarrow \infty} \frac{\boldsymbol{B}}{L} = \sum_i \pi_i \boldsymbol{f}(i) + \sum_{i\neq j} p_{ij} \boldsymbol{g}(i,j).
    \label{eq: average b1}
\end{equation}
The coordinate $(\pi_n,\tau_{ij},J_c)$ spans $(\pi_i,p_{ij})$ linear independently and thus parameterizes the long-term production rates of arbitrary cost-benefit observables.} 
\\

{This is the First-Law-like principle of NEQ. Due to probability conservation, there are only three fundamental flow quantities at steady states: (1) node dwelling without flow, measured by $\pi_n$; (2) symmetric edge flow without net flow, measured by $\tau_{ij}$; and (3) cyclic net flow, measured by $J_c$. This parsing is similar to the landscape-flux decomposition of Markov diffusion processes \cite{graham_covariant_1977,ao_potential_2004,wang_potential_2008,yang_potentials_2021}.}\\

{\textbf{Max Cal derives thermodynamics-like relationships.}}
{Max Cal draws inference---that are minimal \cite{jaynes_minimum_1980,caticha_entropy_2021}, consistent \cite{shore_axiomatic_1980,presse_nonadditive_2013}, and Bayesian-conditioned  \cite{van_campenhout_maximum_1981,csiszar_conditional_1987,chetrite_nonequilibrium_2013,chetrite_nonequilibrium_2015,yang_statistical_2023}---of path probabilities $P^* (\omega)$ for a dynamical model based on constraints on the average of path observables, 
$\bar{\boldsymbol{B}}= \sum_{\omega} P^* (\omega)\boldsymbol{B}(\omega)$, where $P^\ast$ is the posterior model.}
The process $P^*$ maximizes the \textit{path entropy} relative to a prior path probability $P_0$,
\begin{equation}
    \mathfrak{S}_{\text{path}}[P] = -\sum_{\omega} P (\omega) \ln \frac{P (\omega)}{P_0(\omega)},\label{eq: path entropy}
\end{equation}
and satisfies the observable average constraint and normalization of the path probabilities. { 
This is solved by maximizing the Caliber \footnote{In Jaynes 1980 paper where he coined the term \textit{Caliber} \cite{jaynes_minimum_1980}, he refers to the path entropy. Here, we follow \cite{presse_principles_2013} and use Caliber to represent the Legendre transform of path entropy, as the ``free energy-like'' function for nonequilibria.},
\begin{equation}
    \mathfrak{C}[P] = \mathfrak{S}_{\text{path}}[P]+\boldsymbol{\lambda}\cdot\sum_{\omega}P(\omega)~\boldsymbol{B}(\omega), \label{eq: Caliber in P}
\end{equation}
where $\boldsymbol{\lambda}$ is the Lagrange multiplier for the average constraints and normalization is treated implicitly.
}{One gets a Boltzmann-like path distribution with to-be-determined parameters $\boldsymbol{\lambda}$:
\begin{equation}
P_{\boldsymbol{\lambda}}(\omega)=\arg \max_P ~\mathfrak{C}[P]=P_0(\omega)\frac{e^{\boldsymbol{\lambda}\cdot {\boldsymbol{B} (\omega)}}}{\mathcal{Z}(\boldsymbol{\lambda})}
\label{eq: Plambda}
\end{equation} 
where $\mathcal{Z}(\boldsymbol{\lambda}) = \sum_{\omega}P_0(\omega) e^{ \boldsymbol{\lambda}\cdot {\boldsymbol{B}(\omega)}}$ is the dynamical partition function. Substituting Eq. \eqref{eq: Plambda} into \eqref{eq: Caliber in P} shows that Caliber is the \textit{logarithm of the partition function}: $\mathfrak{C}[P_{\boldsymbol{\lambda}}]=\ln \mathcal{Z}(\boldsymbol{\lambda})$.}  
\\

{We now determine the values of $\boldsymbol{\lambda}=\boldsymbol{\lambda}^*$ that gives the observed averaged $\boldsymbol{\bar{B}}$. This follows the same reasoning as applies to EQ (See Sec. II of \cite{SI}).
The derivatives of the Caliber $\mathfrak{C}=\ln \mathcal{Z}(\boldsymbol{\lambda})$ generates the statistical cumulants of the path observable $\boldsymbol{B}$ under the process parameterized by $\boldsymbol{\lambda}$ 
\footnote{For a process parameterized by $\boldsymbol{\lambda}$ with path probability $P_{\boldsymbol{\lambda}}(\omega)=P_0(\omega){[e^{\boldsymbol{\lambda}\cdot {\boldsymbol{B} (\omega)}}]}/{\mathcal{Z}(\boldsymbol{\lambda})}$, its cumulant generating function of the path observable $\boldsymbol{B}(\omega)$ is $G_{\boldsymbol{\lambda}}(\boldsymbol{\alpha})= \ln \sum_{\omega} P_{\boldsymbol{\lambda}}(\omega) e^{\boldsymbol{\alpha}\cdot \boldsymbol{B}(\omega)} = \mathfrak{C}(\boldsymbol{\lambda}+\boldsymbol{\alpha})-\mathfrak{C}(\boldsymbol{\lambda}).$ Therefore, the $n$-th cumulants of $\boldsymbol{B}$ can be computed by the $n$-th derivatives of either functions: $ \nabla^n G_{\boldsymbol{\lambda}}(\boldsymbol{\alpha}) |_{\boldsymbol{\alpha}=\boldsymbol{0}}=\nabla^n \mathfrak{C}(\boldsymbol{\lambda})$.},
so $\boldsymbol{\lambda}^*$ is solved by inverting
\begin{equation}
    \bar{\boldsymbol{B}} = \frac{\partial \mathfrak{C}}{\partial \boldsymbol{\lambda}}\Big|_{\boldsymbol{\lambda} = \boldsymbol{\lambda}^*}.\label{eq: avg = derivative}
\end{equation}
The solution $\boldsymbol{\lambda}^*(\bar{\boldsymbol{B}})$ gives the posterior path probability $P^*=P_{\boldsymbol{\lambda}^*}$, Caliber $\mathfrak{C}^*=\mathfrak{C}[P^*]$, path entropy $\mathfrak{S}^*_{\text{path}}=\mathfrak{S}_{\text{path}}[P^*]$, and their Legendre transform relationship from Eq. \eqref{eq: Caliber in P},
\begin{equation}
\mathfrak{C}^*=\mathfrak{S}_{\text{path}}^*+\boldsymbol{\lambda}^*\cdot \bar{\boldsymbol{B}}.\label{eq: LT}
\end{equation}
These $*$ terms depend implicitly on $\bar{\boldsymbol{B}}$, and the Caliber $\mathfrak{C}^*=\ln \mathcal{Z}(\boldsymbol{\lambda^*})$ has it through $\boldsymbol{\lambda}^*$.
Taking the $\bar{\boldsymbol{B}}$ derivative of Eq. \eqref{eq: LT} shows the path entropic force interpretation of $\boldsymbol{\lambda}^*$ \cite{yang_statistical_2023}:
\begin{subequations} \label{eqs: multipliers as the entropic forces}
\begin{align}
    \frac{\partial \mathfrak{C}^*}{\partial \boldsymbol{\lambda}^*~}   \cdot \frac{\partial \boldsymbol{\lambda}^*}{\partial \bar{\boldsymbol{B}}~~} &= \frac{\partial \mathfrak{S}_{\text{path}}^*}{\partial \bar{\boldsymbol{B}}~~~~~}+\boldsymbol{\lambda}^* + \frac{\partial \boldsymbol{\lambda}^*}{\partial \bar{\boldsymbol{B}}~~}\cdot \bar{\boldsymbol{B}}   \label{eq: entropic force derivation}  \\ \label{eq: 1st line in deriving EP}
    \Rightarrow  \boldsymbol{\lambda}^* &= -\frac{\partial \mathfrak{S}_{\text{path}}^*}{\partial \bar{\boldsymbol{B}}~~~~~}
\end{align}
\end{subequations}
where the cancellation of terms on the two ends follows from Eq. \eqref{eq: avg = derivative}. 
The two equations, Eq. \eqref{eq: avg = derivative} and \eqref{eqs: multipliers as the entropic forces}, together describe the one-to-one Legendre transform between the dual coordinates $\bar{\boldsymbol{B}}$ and $\boldsymbol{\lambda}^*$ for the phase space of the posterior family:
The posterior $P^*$ for different $\bar{\boldsymbol{B}}$ forms a family that can be parameterized by either $\bar{\boldsymbol{B}}$ or their corresponding $\boldsymbol{\lambda}^*$.
}
\\

{\textbf{Large Deviation Theory (LDT) describes the long-time limit.}}
For time-homogeneous processes, the rules our agent follows are the same \textit{at all times}. We thus seek a path probability model $P(i_{0:L})$ that can describe arbitrarily long trajectories $L\rightarrow \infty$.
{
For a Markov jump process with transition rates $k_{ij}$, the path probability can be written down explicitly (See Eq. 21 in \cite{SI}). With respect to a prior with rates $k^{(0)}_{ij}$, the path entropy $\mathfrak{S}_{\text{path}}$ can then be derived and, in the long-term limit, }
is dominated by the {steady-state} path entropy ``rate'', i.e. the path entropy per time step \cite{SI}:  \begin{subequations}\label{eq: path relative entropy rate}
\begin{align}
\mathfrak{s}_{\text{path}}& =\lim_{L\rightarrow \infty} \frac{\mathfrak{S}_{\text{path}}}{L}\\
&= \sum_{i\neq j}p_{ij} - \sum_{i\neq j}\pi_i k^{(0)}_{ij} - \sum_{i\neq j}p_{ij}\ln \frac{p_{ij}}{\pi_i k^{(0)}_{ij}} 
\end{align}
\end{subequations}
where {$\pi_i$ and $p_{ij}$ are the probability distribution and fluxes under the process $P$. The path probability variation for maximizing Caliber in Eq. \eqref{eq: Caliber in P} is reduced to the distribution and flux $(\pi_i,p_{ij})$ variation for maximizing Caliber ``rate'' $\mathfrak{c}=\lim_{L\rightarrow \infty}\mathfrak{C}/L = \mathfrak{s}_{\text{path}}+\boldsymbol{\lambda}\cdot (\sum_i \pi_i \boldsymbol{f}_i+\sum_{i\neq j}p_{ij} \boldsymbol{g}_{ij}).$ with the normalization $\sum_i \pi_i=1$ and stationarity $\sum_{j}(p_{ij}-p_{ji})=0,\forall i$, constraints both handled implicitly.} \\

{
LDT provides the mathematics needed to solve the posterior distribution and fluxes $(\pi_i^*,p^*_{ij})$ and the transition rates $k_{ij}^*=p^*_{ij}/\pi_i^*$ in the long-time limit \cite{chetrite_nonequilibrium_2013,chetrite_nonequilibrium_2015,chetrite_variational_2015}. Essentially, the exponential path probability re-weighting in Eq. \eqref{eq: Plambda} reduces to exponential tilting of the transition rate matrix with eigenvalue-eigenvector calculations. See Sec. III and IV of \cite{SI} for a brief introduction and an example. More examples can be found in \cite{cPRE}. The posterior Markov process family has dual coordinates between average observable rates $\bar{\boldsymbol{b}}$ and forces $\boldsymbol{\lambda}^*$ summarized by the Legendre transform: \begin{equation}
\mathfrak{c}^* = \boldsymbol{\lambda}^*\cdot\bar{\boldsymbol{b}}+\mathfrak{s}^*_{\text{path}}.
\label{eq: LT in rates}
\end{equation}
The derivatives of Caliber rate $\mathfrak{c}(\boldsymbol{\lambda}^*)$ give the asymptotic statistics of the rate observables $\boldsymbol{b}=\boldsymbol{B}/L$ 
\footnote{For the posterior Markov process parameterized by $\boldsymbol{\lambda}$, the scaled cumulant generating function $\text{SCGF}_{\boldsymbol{\lambda}}(\boldsymbol{\alpha})$ of $\boldsymbol{b}$ in LDT is defined as $\text{SCGF}_{\boldsymbol{\lambda}}(\boldsymbol{\alpha})=\lim_{L\rightarrow \infty} (1/L) \ln \sum_\omega P_{\boldsymbol{\lambda}}(\omega)e^{L \boldsymbol{\alpha} \cdot\boldsymbol{b}(\omega)}$.  It is related to the Caliber rate $\mathfrak{c}(\boldsymbol{\lambda})$ by $\text{SCGF}_{\boldsymbol{\lambda}}(\boldsymbol{\alpha})=\mathfrak{c}(\boldsymbol{\lambda}+\boldsymbol{\alpha})-\mathfrak{c}(\boldsymbol{\lambda})$. The $n$-th cumulant of the rate $\boldsymbol{b}$ has the asymptotic form $[\nabla^n \mathfrak{c}(\boldsymbol{\lambda})]/L^{n-1}$ in the limit $L\rightarrow \infty$.}.
} \\


{
In the inference language of Max Cal \cite{evans_rules_2004,presse_principles_2013} or the equivalent LDT language of tilting \cite{maes_canonical_2008,chetrite_nonequilibrium_2013,chetrite_variational_2015,chetrite_nonequilibrium_2015,barato_formal_2015}, Max Cal infers a family of posterior models with thermodynamics-like conjugate relationships. However, the posterior models do not necessarily capture other unconstrained statistics correctly. For example, if only the steady-state distribution $\pi_n$ and net fluxes $J_c$ are constrained as in \cite{maes_canonical_2008}, Max Cal can derive forces conjugated to them. Yet, the inferred traffic change $\tau_{ij}$ is not guaranteed to match the actual change because the model is not constrained by them (see \cite{SI} for details). 
To capture arbitrary changes,  we apply the First Law and constrain all the relevant observables $(\pi_n,\tau_{ij},J_c)$ in Max Cal. This is \textit{not} exploiting Max Cal's power of inference---because the coordinates $(\pi_n,\tau_{ij},J_c)$ already give the rates directly---\textit{but} rather this leverages the ability of Max Cal to compare different processes with nontrivial thermodynamics-like conjugate relationships that have not been developed before. \\
}

{
\textbf{Combining the two laws gives dynamical relationships.}
}For simplicity, we now drop all *'s representing posteriors and denote 
the steady-state averages $(\pi_{n},\tau_{ij},J_c)$ as $\textbf{x}$.  Using $\textbf{x}$ as the set of averages $\bar{\boldsymbol{b}}$ for Max Cal, the posterior transition rates $k_{ij}$ is simply its empirical evaluation from the steady-state distribution and fluxes: \begin{equation}
    k_{ij} =\frac{p_{ij}}{\pi_i} = \frac{\tau_{ij}+J_{ij}(J_c)}{2 \pi_i(\pi_n)}
    \label{eq: k* in pi, tau, J}
\end{equation}
where $J_{ij}(J_c)$ are the edge net flux spanned by the cycle flux $J_c$ and
$\pi_i(\pi_n)=\pi_n$ for $i=n\neq m$ and $=1-\sum_{n'(\neq m)}\pi_{n'}$ for $i=m$. Substituting this back into Eq. \eqref{eq: path relative entropy rate} and choosing a unit-rate random walk $k^{(0)}_{ij}=1$ as the reference \textit{doldrum {process}} with zero forces \footnote{The forces parameterize the differences of a process to the doldrum process. In specific applications, different doldrum process can be chosen as the reference state in the user's convenience. Here, we chose the most symmetric process with baseline activity such that forces characterize how symmetry in dynamics can be broken---like the time reversal symmetry---and how a process can be driven to different activity levels.}, we get a parsing to the path entropy rate:
\begin{align}\label{eq: decomposition of path relative entropy rate}
&\mathfrak{s}_{\text{path}}(\textbf{x}) = \sum_{i}\pi_{i} (\pi_n)\Big[\sum_{j(\neq i)}k_{ij}-N_i\Big] \nonumber\\
&- \frac{1}{2}\sum_{i< j}\tau_{ij}\ln {k_{ij} k_{ji}}- \frac{1}{2} \sum_{c} J_c\ln {\frac{k_{i_0i_1}\cdots k_{i_{\sigma}i_0}}{k_{i_0i_{\sigma}} \cdots k_{i_1 i_0}}}
\end{align}
where $N_i$ is the number of outgoing edges from node $i$ and $i_0 i_1 \cdots i_{\sigma}i_0$ is the sequence of states that specifies the fundamental cycle $c$. Note that the last term is (half negative of) what has been termed ``entropy production rate'' (referring to the production of state entropies, not path entropies) in stochastic thermodynamics \cite{schnakenberg_network_1976,qian_entropy_2016, seifert_stochastic_2018,van_den_broeck_ensemble_2015}. 
We remark here that $k_{ij}$ are functions of all variables in $\textbf{x}$, so the model inference {and comparison} can \textit{not} be divided into parts in general. A variational principle based solely on a part of $\mathfrak{s}_{\text{path}}$, like the minimization or maximization of ``entropic production rate'' \cite{callen_principle_1957,martyushev_maximum_2006},  will fail in general \cite{vellela_stochastic_2008}.\\

Conjugate to the fundamental averages $\textbf{x}$ are the \emph{path entropy forces:} \begin{equation} \label{eq: path entropy force definition}
    {\mathfrak{F}} = -\frac{\partial \mathfrak{s}_{\text{path}}}{\partial {x}}
\end{equation}
where ${x}$ can be any of $\textbf{x}=(\pi_{n},\tau_{ij},J_{c})$.
These are true forces insofar as: (1) they describe the tendencies of $\textbf{x}$ toward particular values; (2) they increase monotonically with the values of the observables \cite{cPRE}, (3) they equal zero when the path entropy is maximal for that component and (4) this definition comports with equilibrium forces such as $1/T = \partial S_\text{state}/\partial U$.  We can frame these forces in terms of the (posterior) transition rates $k_{ij}$ \cite{SI}:\begin{subequations}
\label{force-expression-in-terms-of-the-rates} 
\begin{align}
\mathfrak{F}_{\text{node},n} & =\Big[\sum_{i(\neq m)}k_{mi}-N_m\Big]-\Big[\sum_{j(\neq n)}k_{nj}-N_n\Big]\label{eq:alpha}\\
\mathfrak{F}_{\text{edge},ij} & =  \frac{1}{2} \ln {k_{ij}  k_{ji} }\label{eq:beta}\\
\mathfrak{F}_{\text{cycle},c} & =  \frac{1}{2} \ln {\frac{k_{i_{0}i_{1}} k_{i_{1}i_{2}}\cdots k_{i_{\sigma} i_{0}}}{k_{i_{0}i_{\sigma}} k_{i_{\sigma} i_{\sigma-1}} \cdots k_{i_{1}i_{0}}}}.\label{eq:gamma}
\end{align}
\end{subequations}
An example is as shown in the lower box in Fig. \ref{fig:Parsing}. Note that {these forces are in terms of transition rates \textit{locally} on nodes, edges, or cycles.  The term called cycle affinity in ST is found to be the cycle force, as \textit{derived} from the two laws. }\\

Conjugate to the node distribution $\pi_n$ is the \emph{node force} $\mathfrak{F}_{\text{node}, n}$, with respect to a reference node $m$. These node forces have the units of rates. {Up to constants related to the network topology ($N_n$ and $N_m$), the node force is the escape rate difference compared to the reference node $m$.}
{When a node force $\mathfrak{F}_{\text{node},n}$ is large, the escape rate from node $n$ is small compared to node $m$, leading to large long-term time fraction at node $n$, $\pi_n$}. 
In a truck metaphor, it quantifies how much the truck driver likes to stick in a given node $n$. Recently, it has been shown that $\pi_n$ and the node escape rate has positive correlation for a wide range of processes \cite{calvert_localglobal_2024}. \\

Conjugate to the edge traffic $\tau_{ij}$, is the affinity to the back and forth motion on a given edge, which we call the \textit{edge force} $\mathfrak{F}_{\text{edge},ij}$. As it inherits the (natural) logarithm from the entropy, it has the same units as the entropy rate $\mathfrak{s}_{\text{path}}$. Similarly, the force conjugated to the cycle flux $J_c$ is the \textit{cycle force} $\mathfrak{F}_{\text{cycle},c}$, a notion that has been called \textit{cycle affinity} \cite{schnakenberg_network_1976,hill_free_1989,qian_entropy_2016,yang_bivectorial_2021}. It quantifies the likelihood of a system to finish a forward cycle compared to the backward, and also has the same unit as the entropy. See \cite{jiang_mathematical_2004} for how edge and cycle forces relate to the long-term rate of cycle completion. \\

\begin{figure}
\begin{centering}
\includegraphics[width=0.95\columnwidth]{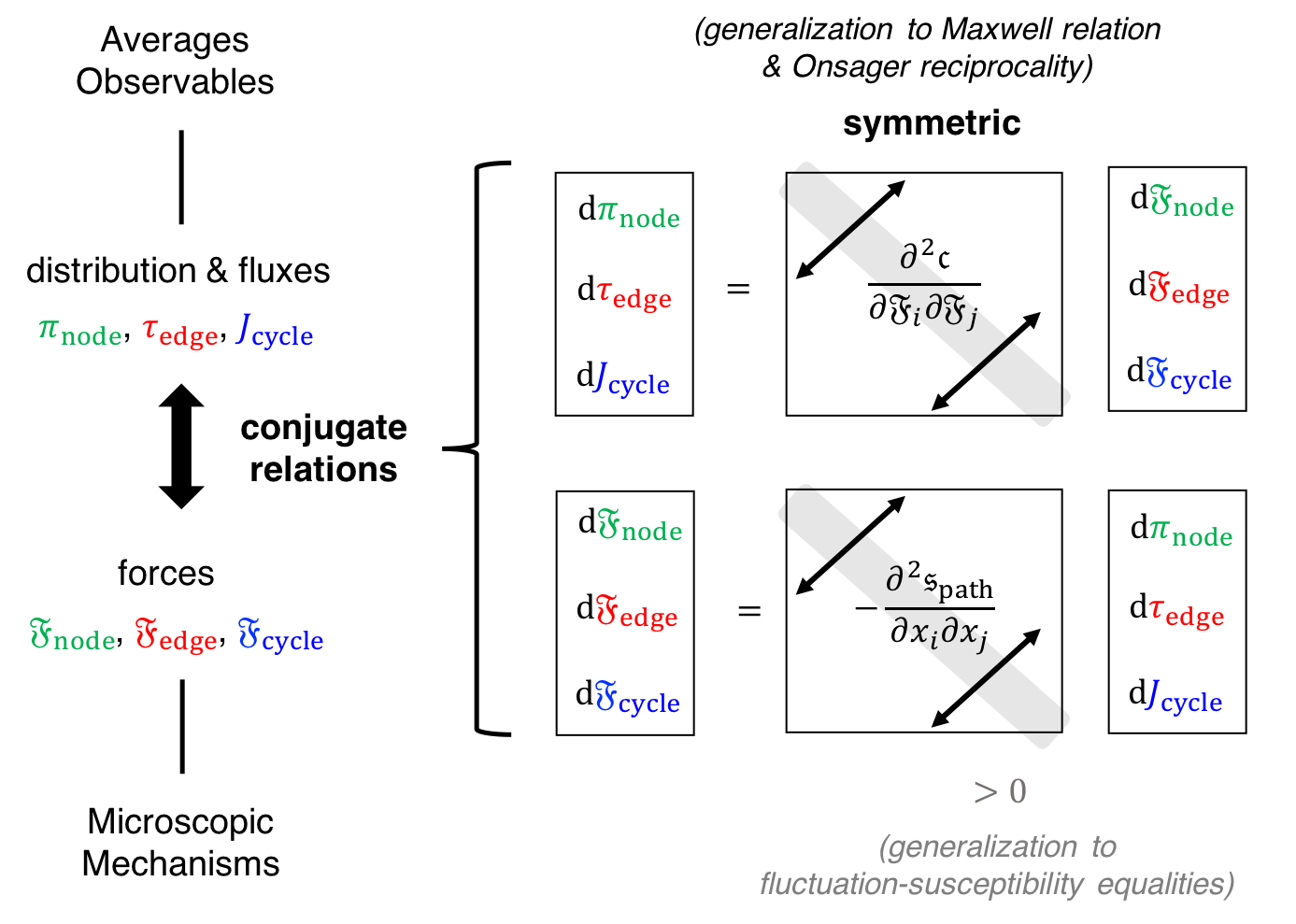}
\par\end{centering}
\caption{\textbf{Conjugate relations connect responses to forces.} On the left, we show that distribution and flux span the average observables $\bar{\boldsymbol{b}}$ and that forces connect to microscopic mechanisms. Conjugated relations are the bridge connecting the two, summarized on the right through susceptibility matrices connecting the infinitesimal changes. The \textit{positivity} of their diagonal components is the generalization to \textit{fluctuation-susceptibility equalities} in EQ whereas the \textit{symmetry} of their off-diagonal components is the generalization to \textit{Maxwell relations} and to \textit{Onsager reciprocality} in EQ and near-EQ systems. \label{fig:parameter-force-observable}}
\end{figure}

\begin{table}
\begin{centering}
\includegraphics[width=0.95\columnwidth]{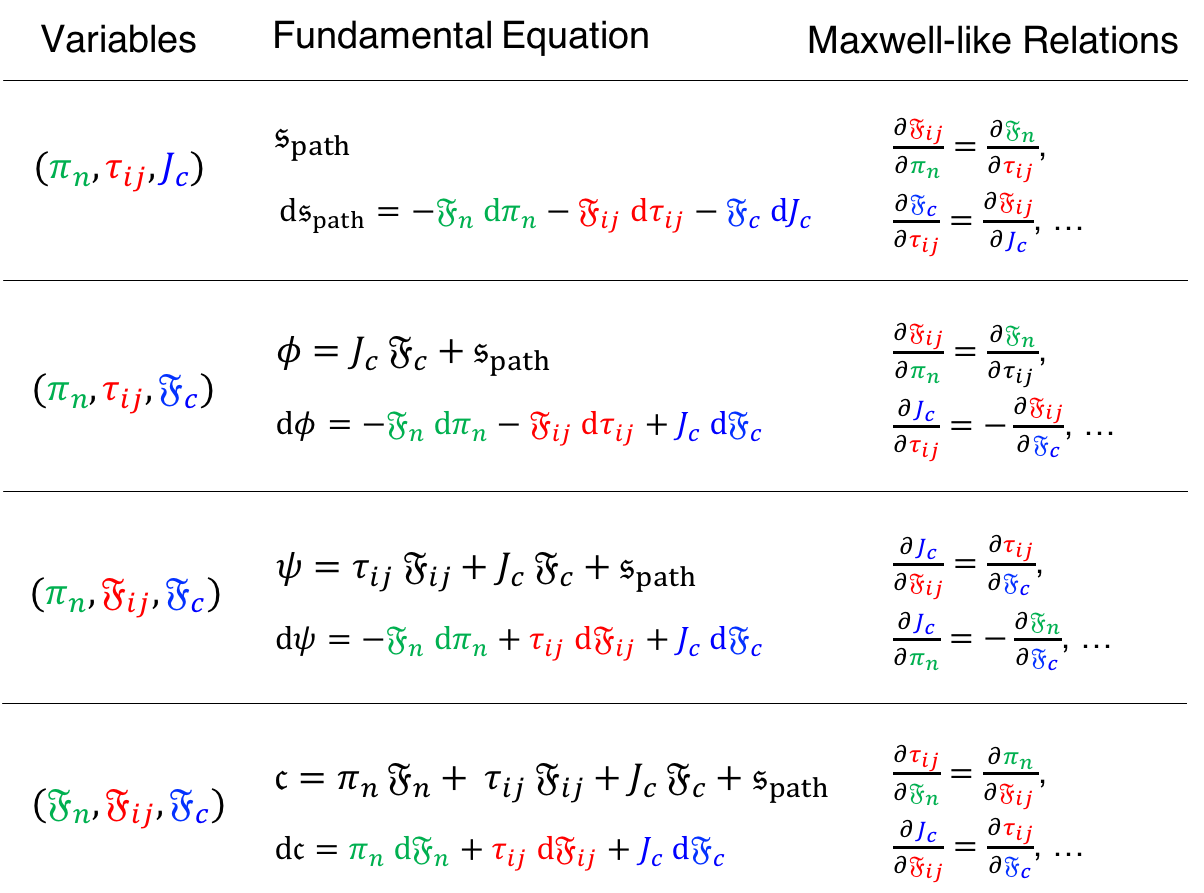}
\par\end{centering}
\caption{\textbf{Maxwell-like Relations in various NEQ dynamic parameterization.} First and Fourth rows show the parameterization in terms of fluxes and forces, respectively. Legendre transform allows definition of various coordinates (ensembles), the associated potentials, and the Maxwell-like relations. The second and third rows are two examples where we swap out one or two (set of) variables with their conjugated variables. Einstein summation convention is used in the middle column for notation simplicity.  \\ 
\label{table: MR table}}
\end{table}

We use $\boldsymbol{\mathfrak{F}}$ as a shorthand for the full set of forces $\left(\mathfrak{F}_{\text{node},n},\mathfrak{F}_{\text{edge},ij},\mathfrak{F}_{\text{cycle},c}\right)$.
{With $\bar{\boldsymbol{b}}=\textbf{x}$ and $\boldsymbol{\lambda}^*=\boldsymbol{\mathfrak{F}}$, the thermodynamic-like conjugate relationship is summarized} 
by the Legendre transform {from Eq. \eqref{eq: LT in rates}}: 
\begin{equation}
\mathfrak{c} = \sum_{\text{node},n}\pi_{n}~ \mathfrak{F}_{n}+\sum_{\text{edge},i j}\tau_{ij} ~\mathfrak{F}_{ij}+\sum_{\text{cycle},c}J_{c}~ \mathfrak{F}_{c} +\mathfrak{s}_{\text{path}},
\label{eq: legendre transform}
\end{equation}  {illustrated}  in Fig. \ref{fig:parameter-force-observable}. We note three important implications here: (1) The \textit{infinitesimal changes} in $\textbf{x}$ and $\boldsymbol{\mathfrak{F}}$ are coupled by susceptibility matrices that have \textit{positive diagonal} components. This is the NEQ generalization of \textit{fluctuation-susceptibility relation}, fluctuation here{---computed from the second derivative of $\mathfrak{c}(\boldsymbol{\mathfrak{F}})$---is the asymptotic variance of distribution and flux variables with averages $\textbf{x}$}; (2) The susceptibility matrix has \textit{symmetric off-diagonal} terms. This follows directly from Eq. \eqref{eq: path entropy force definition} and {Eq. \eqref{eq: avg = derivative} with $(\bar{\boldsymbol{B}},\mathfrak{C},\boldsymbol{\lambda})$} replaced by {$(\textbf{x},\mathfrak{c},\boldsymbol{\mathfrak{F}})$}. It is the direct generalization to \textit{Maxwell relations} and \textit{Onsager reciprocality} in EQ and near-EQ systems; (3) The \textit{Legendre transform} between $(-\mathfrak{s}_{\text{path}})$ and $\mathfrak{c}$ allows defining \textit{alternative coordinates (NEQ ensembles)} and deriving the Maxwell relations in them, \textit{e.g.} see Table \ref{table: MR table}.
We present further illustration of these conjugated relations and applications in \cite{cPRE}. \\

{\textbf{Summary} 
EQ thermodynamics utilizes observable coordinates such as $(U, V, N)$, and derives their conjugate forces as $(1/T, p/T, -\mu/T)$.  We show here that for general NEQ, the relevant coordinates are $\textbf{x}=(\pi_\text{node},\tau_\text{edge},J_\text{cycle})$ and the path-entropy forces are $\boldsymbol{\mathfrak{F}}=(\mathfrak{F}_\text{node},\mathfrak{F}_\text{edge},\mathfrak{F}_\text{cycle})$.  We also derive the NEQ Legendre Transforms, the fluctuation-susceptibility equalities, and the Maxwell-Onsager-like symmetry relations.} \\

\begin{acknowledgments}
We thank the Laufer center for support, and {the anonymous reviewers}, Charles Kocher, Olga Movilla Miangolarra, Jonathan Pachter, Rostam Razban, Yuhai Tu, Lakshmanji Verma, and Jin Wang for insightful feedback. Y.-J. expresses his deepest gratitude to Hong Qian for guiding him in learning the essential theoretical pieces used in this work. 
\end{acknowledgments}

\bibliography{main-PRL-rv1.bbl}

\end{document}